\documentclass[12pt,journal]{IEEEtran}
\usepackage{cite}

\usepackage{graphicx}
\usepackage{subfigure}
\usepackage{epstopdf}
\usepackage{multicol}
\usepackage{algorithm}
\usepackage{algorithmic}
\usepackage{amsthm,amsmath,amsfonts}
\usepackage{mathrsfs}
\usepackage{courier}

\usepackage{bm}
\usepackage{subfigure}

\usepackage{color}

\def\BibTeX{{\rm B\kern-.05em{\sc i\kern-.025em b}\kern-.08em
    T\kern-.1667em\lower.7ex\hbox{E}\kern-.125emX}}

\begin{document}
%
\title{Networking of Internet of UAVs: Challenges and Intelligent Approaches}
%
%
%
\author{Peng Yang, Xianbin Cao, Tony Q. S. Quek, and Dapeng Oliver Wu
}

\maketitle

\begin{abstract}
Internet of unmanned aerial vehicle (I-UAV) networks promise to accomplish sensing and transmission tasks quickly, robustly, and cost-efficiently via effective cooperation among UAVs. 
To achieve the promising benefits, the crucial I-UAV networking issue should be tackled. 
This article argues that I-UAV networking can be classified into three categories, quality-of-service (QoS) driven networking, quality-of-experience (QoE) driven networking, and situation aware networking.
Each category of networking poses emerging challenges which have severe effects on the safe and efficient accomplishment of I-UAV missions. 
This article elaborately analyzes these challenges and expounds on the corresponding intelligent approaches to tackle the I-UAV networking issue. 
Besides, considering the uplifting effect of extending the scalability of I-UAV networks through cooperating with high altitude platforms (HAPs), this article gives an overview of the integrated HAP and I-UAV networks and presents the corresponding networking challenges and intelligent approaches. 

\end{abstract}


%
\IEEEpeerreviewmaketitle

\section{Introduction}
\IEEEPARstart{T}{he} sixth generation (6G) wireless communication networks are desired to provide ubiquitous and seamless geographical communication coverage to meet diverse use cases in many scenarios including villages and motorways \cite{DBLP:journals/chinaf/YouWHGZWHZJWZSW21}. Obviously, it is difficult to achieve the above-mentioned ambitious goal with terrestrial networks (TNs) alone. For TNs, they are vulnerable to natural disasters, severe ground disruption. As a result, ground (mobile) users will experience communication interruptions. Since the interruptions are either temporary or unexpected, it will be timely infeasible to construct TNs to recover communications. 
In this case, resorting to the assistance of non-terrestrial networks (NTNs) is a promising selection in terms of cost-effectively implementing the above goal. Actually, in 6G era, the integration of TNs and NTNs is a global consensus, and the demonstration on the integration is initiated in many counties.

NTNs are composed of many heterogeneous interconnected flying platforms deployed at different altitudes ranging from tens of meters to tens of thousands of kilometers, including satellites, high altitude platforms (HAPs), and unmanned aerial vehicles (UAVs). 
These platforms have advantages and disadvantages concerning such aspects as cost, persistence, responsiveness, vulnerability, footprint, and overflight \cite{DBLP:journals/jsac/CaoYAXWY18}.
Owing to the unique advantages in terms of cost, flexibility, responsiveness, and communication latency, UAV-assisted wireless communications have received extensive attention from both academia and industry. UAVs mounted with diverse devices have also been applied to accommodate some typical use case demand, e.g., UAV-base station (BS) for ubiquitous coverage, UAV-relay for distant users' connection \cite{DBLP:journals/cm/ZengZL16}. 

Despite the many promising advantages, the design of a single UAV network faces many tricky challenges, for example, unreliable communication link, weak survivability, small footprint, long mission completion time, size, weight, and power (SWAP) constraints, and so on. 
To meet these challenges when accomplishing a mission, one needs to construct an architecture of Internet of UAVs (I-UAVs) involving many cooperative UAVs. 
Particularly, through the I-UAVs, the mission completion time can be decreased, the scalability (i.e., coverage area) can be extended, the survivability can be increased, the sensing ability can be enhanced, and the detectability can be decreased \cite{DBLP:journals/adhoc/BekmezciST13}. 

Nevertheless, to achieve so many promising benefits, one of the key issues of I-UAVs must be solved, i.e., the networking of I-UAVs. The I-UAV networking is defined as the deployment of many cooperative UAVs to accomplish the sensing and transmission tasks safely and efficiently. From the viewpoint of diverse purposes, the I-UAV networking can be classified into three categories, that is, quality-of-service (QoS) driven I-UAV networking, quality-of-experience (QoE) driven I-UAV networking, and situation aware I-UAV networking. Further, conducting the I-UAV networking poses many challenges which have severe effects on the safe and efficient accomplishment of I-UAV missions. 


The primary task of the QoS-driven I-UAV networking is to guarantee the QoS of ground users, which are usually characterized by users' achievable data rates. 
Because of the dynamic stochastic deployment environment and user mobility, the problem of I-UAV networking has to be formulated as a sequence-decision problem, subject to multiple physical and topology constraints (e.g., UAV energy consumption, UAV service ability, and outage probability). This problem can be confirmed to be non-deterministic polynomial (NP) hard, which is difficult to be solved by some conventional optimization approaches. 

As reported by Cisco, mobile video traffic is expected to occupy about 79\% of global mobile data traffic by 2022 \cite{Cisco_index}. Besides, 80\% of the mobile video traffic belongs to hotspot contents (e.g., FIFA World Cup, American Super Bowl), the coverage of which is one of the typical use cases of deploying I-UAVs. 
Therefore, performing I-UAV networking to deliver video streams for ground users will be a key mission of I-UAV networks. However, how to deploy I-UAVs to guarantee the QoE requirements of ground users is challenging. First, QoE is an application layer indicator in the multimedia transmission field, which is particularly subjective. Thus, finding an appropriate QoE model which can exactly correlate it with low-layer and controllable resources is non-trivial. 
Second, reliable and low-latency propagation links are desired for video transmission. Yet, the intrinsic characteristic of random channel fluctuations in I-UAV networks may result in playback buffer starvations and then video freezes. 

Additionally, UAVs are deployed in complex and shared three-dimensional (3D) airspace. Situation information must be considered when performing the I-UAV networking to ensure the safe flight of UAVs and efficient video stream transmission and so on. However, the investigation on the situation aware I-UAV networking is difficult. First, the dynamic stochastic 3D environment where various unexpected scenes may happen in burst will threat the computing capability and safety of I-UAV networks. In this case, how to correctly sense and fuse the local situation built by a UAV is challenging. Second, there are many UAVs in I-UAV networks following with many sensors, great sensing range and large amount of situation information, then how to share location situation and construct a global situation field in I-UAV networks is difficult. 
Third, owing to the existence of high-dimensional multi-modality situation, conventional optimization approaches will be not sufficient to solve the I-UAV networking problem.

Summarily, the goal of this article is to present an overview of the I-UAV networking. The basic architecture of I-UAV network, intelligent modeling of channel gain of I-UAV network, emerging challenges and intelligent approaches of I-UAV networking, as well as potentials and challenges of the networking of integrated HAP and I-UAV networks are presented. 


\begin{figure*}[!t]
\centering
\fbox{\includegraphics[width=4.3in]{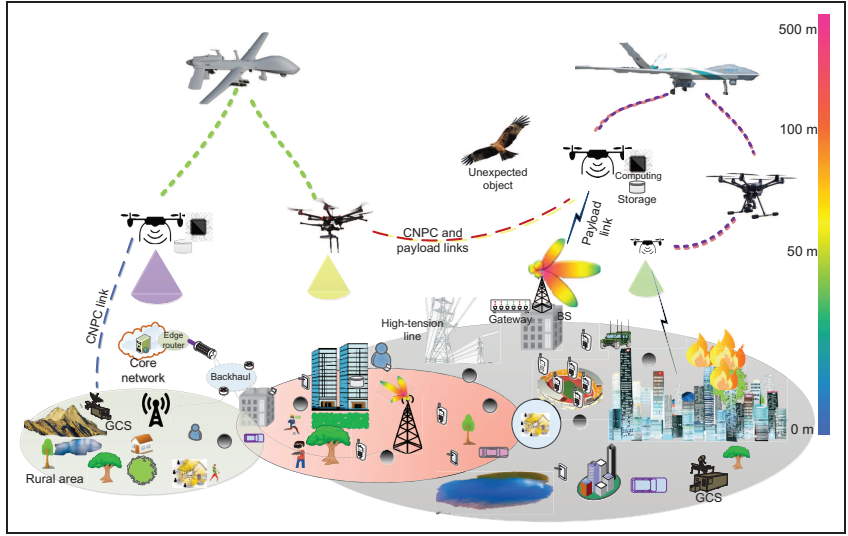}}
\caption{Basic architecture of I-UAV networks.}
\label{fig_UAV_networking}
\end{figure*}

\section{Basic Architecture of I-UAV networks}
As shown in Figure \ref{fig_UAV_networking}, I-UAV networks consist of multiple cooperative fixed-wing and rotary-wing UAVs which are flying at altitudes of tens to hundreds of meters. 
Depending on the roles of UAVs in the network, the network can provide different services. For instance, the UAVs in the network can act as flying base stations (BSs) with BSs being mounted on them. The I-UAV networks can then be applied in several emergency communication scenarios, such as flash crowd traffic offloading from a congested ground BS and temporary events (e.g., gathering and sport event). Assisted by I-UAV networks, the communication coverage and capacity of a ground cellular network can also be quickly enhanced. 
Besides, the UAVs in the network can act as flying access points (or relays) when fronthaul/backhaul hubs are mounted on them. The I-UAV networks can then form an airborne fronthaul/backhaul hub network, which collects fronthaul/backhaul traffic from ground BSs and forward aggregated traffic back to a ground gateway. 

Except for mounting a BS or a hub, each UAV in the network can be equipped with multiple types of sensors and computing and communication modules for stabilization, navigation, positioning, sensing, communication and so on. 
The sensors include three-axis accelerometer, three-axis gyroscope, magnetometer, barometer, GPS, distance sensor, electro-optical pod and so on. 
For the computing module, it is responsible for analyzing and fusing sensed data, e.g., target detection and recognition. 
The communication module will support two types of communication links, i.e., control and non-payload communication (CNPC) link and payload communication link. 

The CNPC link is established to ensure the safe operation and efficient control of I-UAV networks. To this aim, dedicated frequency band has been allocated for the UAV CNPC link, i.e., 960–977 MHz at the L-band and 5030–5091 MHz at the C-band \cite{DBLP:journals/vtm/MatolakS15}.
As shown in Figure \ref{fig_UAV_networking}, in the I-UAV networks, a CNPC link must be established between a UAV and a ground control station (GCS). This is because the human intervention of the UAV in case of an emergency is clearly stipulated by the law. Meanwhile, CNPC links need to be maintained among UAVs for exchanging partial situation information (e.g., safety and control information).

For the payload link, it is established for enabling the mission-related (or payload) communications of I-UAV networks. To support the payload transmission, the frequency band 2.4-2.4835 GHz at the S-band and 5.725-5.85 GHz at the C-band are allocated for the payload link. 
As shown in Figure \ref{fig_UAV_networking}, the payload link will be established between a UAV and a ground BS or a gateway for traffic offloading and delivery. The payload link will also be maintained by UAVs for data transmission (e.g., video stream) and local situation information (e.g., detected or tracked targets) exchanging. 



\section{Intelligent modeling of Channel gain of I-UAV networks}
I-UAV networks consist of three types of communication channels, i.e., UAV-to-UAV (UtU) channel, UAV-to-ground user (UtG) channel, and UAV-to-base station (UtB) channel, as shown in Figure 1. 
Compared to well-studied ground communication channels, communication channels of I-UAV networks exist some unique characteristics, and one needs to explore intelligent approaches to model them accurately. 

\subsection{UAV-to-UAV channel}
Owing to the high deployment altitude and flexible mobility in 3D airspace, the signal among UAVs is generally considered to propagate via line-of-sight (LoS) links. Correspondingly, many UAV-assisted communication works leverage LoS propagation models (e.g., Friis transmission equation) to characterize the UtU channel. Except for receiving LoS signal components, however, a UAV will receive more complicated non-line-of-sight (NLoS) signal components, for example, reflected signals from terrains, high-rise buildings, scattered signals from the atmosphere when the UAV works at the S-band. Summarily, a UAV may receive both LoS and complicated NLoS signal components, and all of the received signal components should be considered when modeling the UtU channel such that more accurate channel characteristics can be captured. 
Additionally, UtU channel may experience the Doppler frequency shift, the amount of which is closely related to UAVs' relative velocities. 
Nevertheless, the consideration of complicated NLoS components and the Doppler frequency shift as well will significantly hinder the theoretical derivation of a closed-form expression of the UtU channel and may result in a channel expression intertwined with many parameters. In this case, the subsequent problem of I-UAV networking may become theoretically intractable. 

To tackle this issue, deep neural networks (DNNs) can be explored. DNNs have the well-known powerful nonlinear approximation ability. The universal approximation theorem asserts that a neural network (NN) with one hidden layer and enough hidden neurons is sufficient to approximate a continuous mapping when proper activation functions are selected \cite{DBLP:books/daglib/0029547}. Further, DNNs with deeper layers can be leveraged to approximate more complex continuous mappings. As a result, by training DNNs using measured channel coefficients, the complicated channel expression can be approximated by a relative simple one as a function of UAVs' relative velocities and the distance between two UAVs. 

Besides, most UAV-assisted communication works solved a specific communication problem based on a key assumption of a single transceiver antenna. Although UAVs are energy constrained and sensitive, it is possible to be equipped with several antennas to significantly alleviate the impact of channel fading and improve the transmit data rate via a multi-antenna technique. Certainly, the effective application of multiple input and multiple output (MIMO) technique in I-UAV networks needs to tackle many challenging issues (especially, dynamic channel estimation and tracking). To this aim, machine learning (ML) and artificial intelligent (AI) approaches can be explored to predict and approximate channels and incorporate the UAV-related parameters (e.g., moving direction, deployment altitude, and antenna orientation) into the dynamic UtU channel.

\subsection{UAV-to-ground user channel}
The investigation on the modeling of UtG channels is one of the most hot research topics in UAV communications. Owing to the surrounding complicated reflection, diffraction, and scattering environments, ground users will receive signals from a large number of propagation paths. 
Motivated by this observation, geometry-based stochastic channel model was developed by considering the signal scatter and reflection on some standard geometric shapes (e.g., cylinder, sphere, ellipsoid). Yet, the mathematical expression of the channel gain or channel impulse response (CIR) of this type of channel model is rather complicated.
To simplify the theoretical expression of the UtG channel gain, statistical analysis methods (e.g., fitting and estimation) were explored. Specifically, given measured channel coefficients, statistical analysis methods will derive a channel closed-form function to approximate or average the coefficients. Nevertheless, the statistical channel model is closely related to the actual UAV deployment environment, and one statistical channel model is not fit for all UAV deployment environments. 
It is known that the radio propagation environment is time-varying and atmosphere will affect the attenuation degree of radio propagation. 
Therefore, the statistical channel model cannot effectively reflect the attenuation characteristic of the time-varying UtG channel. 
Using ML/AI approaches to approximate the UtG channel dynamically will be a promising solution for the UtG channel modeling. 
For example, considering an urban area of size $1 \times 1$ km$^2$ where buildings are generated by the International Telecommunication Union (ITU) recommended local building model with statistical parameters $\alpha = 0.3$, $\beta = 300$ buildings/km$^2$, $\gamma$ being modeled as a Rayleigh distribution with the mean value $\delta = 30$ m \cite{Propagation2012ITU}.
Generate the UtG path-loss using the 3GPP specification \# 36.777 path-loss model for urban Macro given in Table B-2 of \cite{Technical20173GPP}. The small-scale fading coefficient is added assuming Rayleigh fading for the NLoS case and Rician fading with 15 dB Rician factor for the LoS case, where the presence/absence of an LoS link between a UAV and a mobile user can be determined based on the building realization. 
In this way, the actual UtG channel gain coefficients can be obtained. To approximate them, a UAV can construct and train a unique DNN to estimate UtG channel gain coefficients between it and all mobile users. The input of the DNN includes locations of the UAV and the user connecting to it. 
Then, one can online train the DNN using periodically measured channel coefficients and leverage the trained DNN to predict UtG channel coefficients \cite{DBLP:journals/jsac/YangXGQCC21}. 
\begin{figure*}[!t]
	\centering
	\fbox{\subfigure[]{\includegraphics[width=3.3 in]{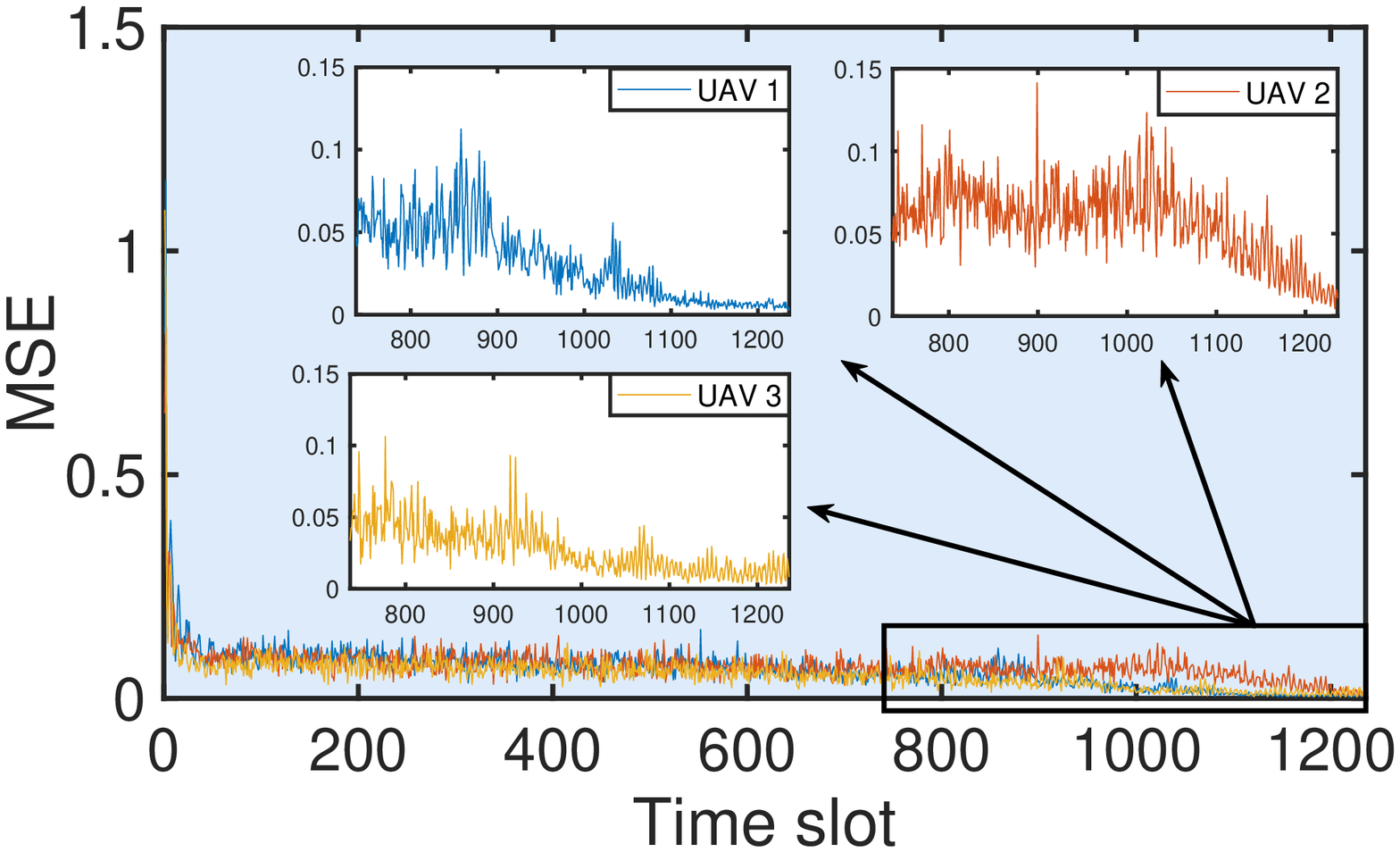}
			\label{fig_dnn_estimation}}
		\hspace{4mm}
		\subfigure[]{\includegraphics[width=3.3 in]{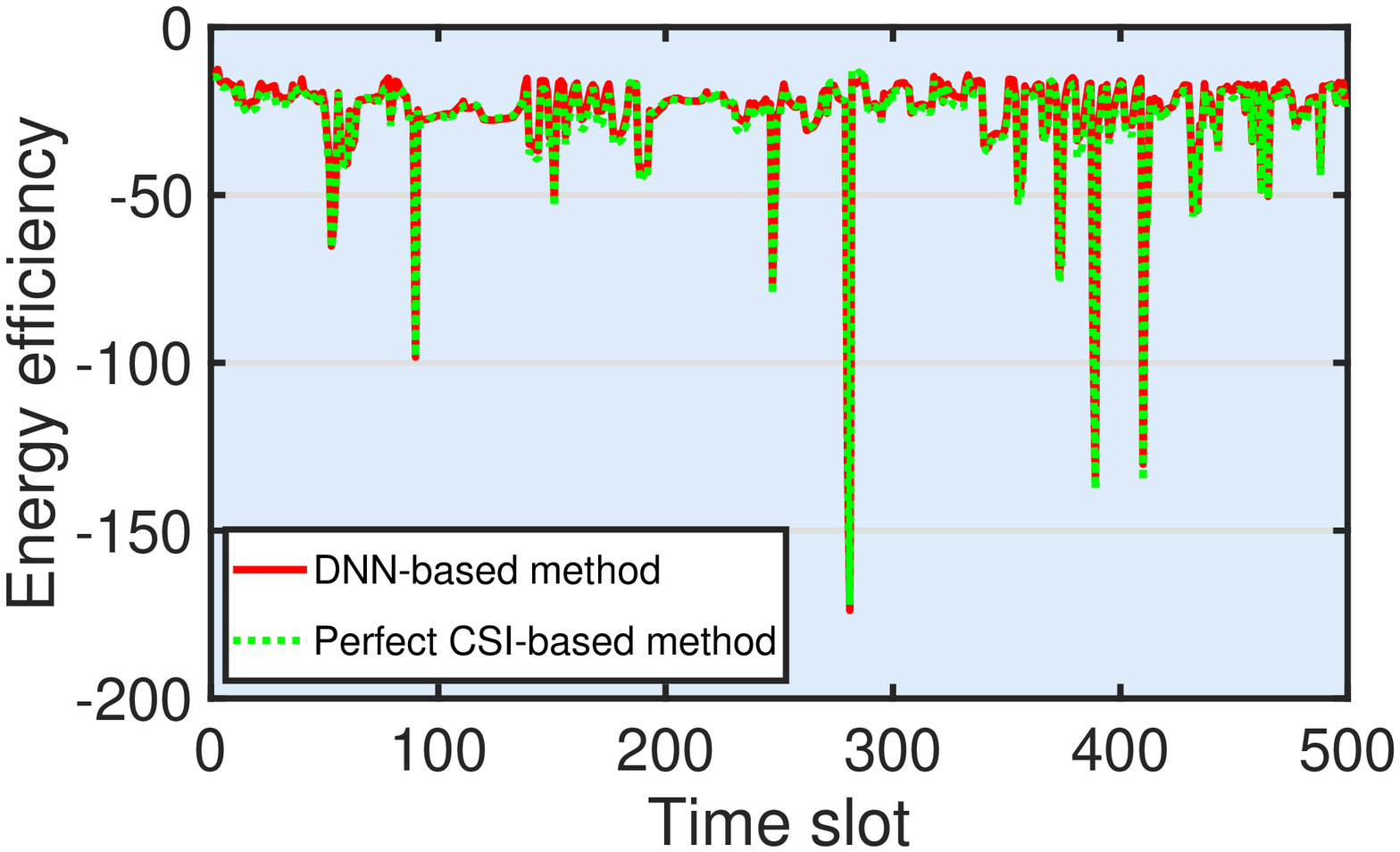}}
		\label{fig_instant_ee}}
	\caption{a) Mean squared error (MSE) of DNNs for UtG channel gain coefficient estimation versus time slot; b) comparison of the obtained instantaneous energy efficiency by the proposed DNN-based method and perfect CSI-based method.}
	\label{fig_MSE_EE}
\end{figure*}
This is illustrated in Figure \ref{fig_dnn_estimation} with the pre-training (or offline training) duration being $736$ time slots and the online training duration being $500$ time slots. For each DNN, its $1^{\rm st}$ hidden layer has 512 neurons and the $2^{\rm nd}$ hidden layer has 256 neurons. Three UAVs are deployed at a constant flight altitude $H = 50$ m, and the system bandwidth is $10$ MHz. 
It is observed from Figure \ref{fig_MSE_EE} that the estimation error is great initially, but it quickly decreases with the increase of time slots as more experience is accumulated. After a number of time slots, DNNs for UtG channel gain coefficient estimation can converge. 
Although actual UtG channel gain coefficients vary fast due to the movement of users and UAVs, the proposed DNN-based method can achieve good estimation results. For example, during the online training period, the loss values of DNNs for UtG channel gain coefficient estimation reach an order of less than 1.5e-1. The obtained energy efficiency of the proposed method is close to that of the perfect channel state information (CSI) based method. 

Considering the important role of UAVs in 6G and the critical impact of UtG channel on I-UAV networking, more attention is desirable to be paid on the intelligent UtG channel modeling. 


\subsection{UAV-to-base station channel}
In many researches on UAV-assisted ground cellular networks, UtG channel gain models are directly adopted to model the UtB channel. This is inappropriate as large antenna arrays can be mounted on a ground BS while ground users are usually equipped with one or several antennas. As a result, the channel characteristics of UtB channels are significantly different from that of UtG channels. 
Additionally, for a BS equipped with full-dimensional large arrays, it can leverage the massive MIMO technique to mitigate interference and significantly enhance the network spectral efficiency. 

To benefit from this advanced technology, the most important issue is the dynamic UtB channel estimation and tracking in 3D airspace. Owing to the highly dynamic movement of UAVs, UtB channel phase varies rapidly over time, and the azimuth and elevation angles of UtB channels change quickly. These characteristics will greatly affect the timeliness and accuracy of the UtB channel estimation and tracking. 
Besides, the UAV's vibration will impact the accuracy of the direction of arrival (DoA) estimation. How to compensate the DoA estimation error caused by UAV's vibration is a crucial while under-studied research topic. 
To estimate the time-varying UAV azimuth and elevation angles, some conventional methods was leveraged, e.g., angular speed-based and Kalman filter-based channel prediction methods. 
Yet, to further improve the accuracy and robustness of channel tracking for UAVs, ML/AI approaches should be explored. For example, one can first simulate the law of UAV machine vibration using ML/AI approaches and then correct/compensate the DoA estimation error with the simulated results. 


\section{Intelligent I-UAV networking}
This section presents three types of I-UAV networking, i.e., QoS-driven I-UAV networking, QoE-driven I-UAV networking, and situation aware I-UAV networking. The emerging challenges of investigating these types of networking are elaborately analyzed, and the corresponding intelligent networking approaches are expounded. 

\subsection{QoS-driven Intelligent I-UAV networking}
To satisfy the QoS requirement of ground users is one of the most significant goals of I-UAV networking. Generally, the QoS of a user is characterized as its achievable data rate. 
During the past five years, numerous works related to the I-UAV networking in two-dimensional (2D) airspace were published. Most of them propose to formulated the I-UAV networking problem as a mathematical programming problem. The primary objective of the formulated problem is to accommodate users' QoS, subject to computing, networking and storage resource constraints.
To this aim, some different types of resources such as UAV caching, bandwidth allocation, UtG association, UAV path planning, UAV and user transmit power are optimized using conventional optimization methods. 
The formulated mathematical programming problem can be classified into a sequential-decision problem, where the problem at each sequence is a mixed integer nonlinear programming (MINLP) problem. 
Fortunately, the Lyapunov technique can be explored to decompose the sequential-decision problem, and iterative and approximation schemes can then be leveraged to solve the decomposed problem \cite{DBLP:journals/jsac/YangXGQCC21}.

In practical application scenarios, UAVs can flexibly fly in 3D airspace; thus, increasing attention has been paid to study the case of I-UAV networking in 3D airspace. Nevertheless, the formulated I-UAV networking problem in 3D airspace is much more challenging than that in 2D airspace. First, higher-dimensional decision variables need to be optimized in a 3D I-UAV networking problem. Second, 3D locations of UAVs are complicatedly coupled in the UtG channel model, which poses a challenge to the theoretical tractability. In this case, it is difficult (if it is possible) to solve the formulated problem using conventional optimization methods. Reinforcement learning (RL), which can effectively eliminate the risk of combinatorial explosion, has been demonstrated as an efficient way of handling complex control problems in continuous and high dimensional state spaces. As a result, many researchers turn to leverage RL methods to solve the 3D I-UAV networking problem. 

\begin{figure}[!t]
	\centering
	\fbox{\includegraphics[width=3.4in]{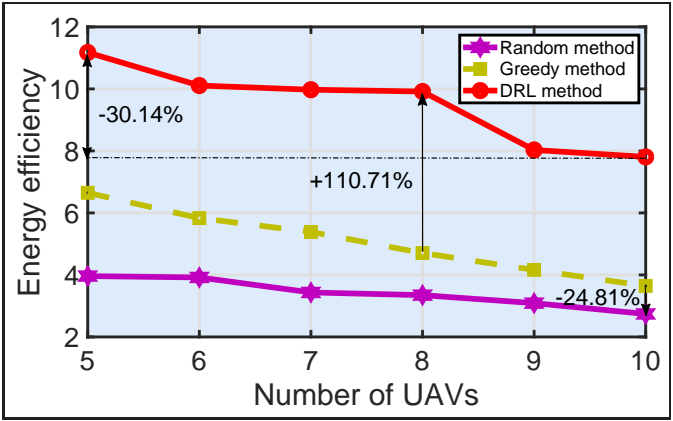}}
	\caption{Comparison of the achieved energy efficiency versus number of UAVs by the proposed DRL method, a method of randomly selecting UAVs' movement, and a greedy method of selecting UAVs' movement that maximize the objective value.}
	\label{fig_EE_UAV_NUM}
\end{figure}

For example, in \cite{DBLP:journals/tvt/YangCXDXW19}, we consider a communication scenario where I-UAV networks with $J$ UAVs are centrally controlled to continuously fly in 3D airspace to accomplish a mission of providing energy-efficient and fair data delivery services for a number of $N$ quasi-stationary users. These users are uniformly distributed in a geographical area of $2.5 \times 2.5$ km$^2$. At each time slot, a UAV can connect to at most one user, and a user can be served by at most one UAV. The connection between a UAV-user pair is considered to be established only if the user's QoS requirement is satisfied. Then, the continuous movement control problem of I-UAV networks can be formulated as a sequential-decision problem aiming at maximizing the energy-efficient and fair communication coverage, subject to constraints on users' QoS requirements, UAVs' flight airspace, and the connectivity of I-UAV networks. To solve this problem, a new deep reinforcement learning (DRL) method is developed, where all the actor and critic networks are two-layer fully-connected feedforward NNs. 
Besides, in this method, the action space consists of UAVs' moving distances, pitch and yaw angles. The actions resulting in the airspace boundary and the network connectivity constraints will be penalized. 
Figure \ref{fig_EE_UAV_NUM}  illustrates the tendency of the obtained energy efficiency and Jain’s fairness index with a constant UAV transmit power $P_D = 24$ dBm and the number of users $N = 100$. 
The $1^{\rm st}$ and $2^{\rm nd}$ hidden layers have $400$ and $300$ neurons, respectively. The minimum and maximum allowable UAV deployment altitudes are $100$ m and $800$ m, respectively. 
It is observed from Figure \ref{fig_EE_UAV_NUM} that the proposed DRL method can effectively solve the UAV movement control problem although more UAVs will make the problem more difficult to be tackled. 
The proposed method can also significantly enhance the fairness and improve the energy efficiency of the communication coverage of I-UAV networks. 




\subsection{QoE-driven Intelligent I-UAV networking}
It is widely considered that providing high quality mobile video services for ground users is one of the most significant goals of deploying I-UAV networks. The quality of received videos by ground users is quantified by QoE of users. From the perspective of video transmission, the QoE of a user is defined as its subjective measurement towards perceived video streams, which is primarily affected by network status (e.g., bandwidth, latency, and throughput) and content configuration factors (e.g., coding, resolution, and sampling rate) \cite{DBLP:journals/comsur/ZhaoLC17}. 

\begin{figure}[!t]
	\centering
	\fbox{\includegraphics[width=3.4in]{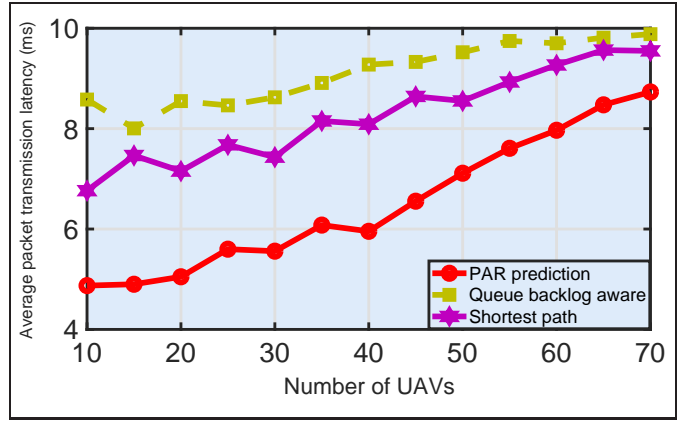}}
	\caption{Comparison of the obtained average packet transmission latency versus the number of UAVs by the proposed PAR prediction method, a method with the queue backlog value being a routing metric, and the shortest path routing method.}
	\label{fig_latency_UAV_NUM}
\end{figure}

On one hand, owing to the high flexibility and controllable mobility, LoS UtG channels can be easily established for I-UAV networks to improve the quality of mobile video services. On the other hand, the mobility of UAVs will lead to time-varying I-UAV network status. In this case, it is difficult to accommodate QoE requirements of users. 
For example, unreliable UtG links will dramatically degrade video demodulation quality, and a single packet loss may result in video freezes for several seconds. 
Therefore, it is urgently required to solve the highly challenging I-UAV networking problem to guarantee the QoE requirements of users. 
To tackle this problem, ML/AI approaches can be explored. For instance, one can leverage ML/AI approaches (e.g., echo state network (ESN), long short-term memory (LSTM)) to predict stochastic packet arrival processes. With the predicted results as one of routing metrics, novel UAV routing protocols can capture the packet accumulation situation in each UAV and then balance network load and relieve network congestion. 
For example, let us imagine a communication scenario where I-UAV networks with $J$ UAVs are deployed to sense a disaster area and transmit sensed video streams back to a GCS for disaster analysis. To alleviate network congestion, a typical UAV $j$ will choose another UAV $k$ in its communication range as its video delivering node (or called next hop node). Whether UAV $k$ will be selected or not is determined by its queue backlog length $l_k$, the transmission latency $d_k$ between it and UAV $j$, and its minimum hop-counts $h_k$ towards the GCS. Since the packet arrival rate (PAR) determines the queue backlog length and directly reflects packet arrival processes, an LSTM approach is explored to predict future PARs and the corresponding queue backlog length according to historically observed PARs. Then, the UAV $k$ with the minimum weighted sum of normalized $l_k$, $d_k$ and $h_k$ will be UAV $j$'s next-hop node. Meanwhile, if UAV $k$ does not acknowledge (ACK) packets from UAV $j$ within $T^{\rm th}$ ms, UAV $j$ will re-start the above-mentioned next-hop node selection mechanism. Obviously, compared to routing methods without PAR prediction, the novel routing method will significantly reduce the packet backlog.
This is illustrated in Figure \ref{fig_latency_UAV_NUM} with a constant UAV communication radius $r = 10$ m and the maximum ACK waiting time $T^{\rm th} = 10$ ms. Figure \ref{fig_latency_UAV_NUM} shows that the novel UAV routing protocol outperforms the shortest path routing protocol and queue backlog aware routing protocol in decreasing the average packet transmission latency. 

Additionally, the video transmission has the stringent low transmission latency and high reliability requirement. Owing to the movement of users and the time cost of configuring I-UAV networks (including network topology and resources), ML/AI approaches can also be explored to realize the proactive I-UAV networking such that transmission latency is reduced. 
To guarantee the transmission reliability while reducing transmission latency, the fountain code technique is recommended to be adopted in the I-UAV networking. Besides, even though the fountain code technique is explored, dramatically changing network status (e.g., link capacity) will lead to the video timeout and the lack of playable videos in the video playback buffer and then the appearance of video freezes. With the network status being proactively obtained, the dynamic adaptive streaming over HTTP (DASH) technique can further be explored to change source coding rates to adapt to the network status. 
Certainly, defining good behavior of a video service is particularly subjective, and many factors will affect the quality of video services, for example, the current video quality level, the oscillation in quality levels during the video playout and buffer starvations, leading to video freezes \cite{DBLP:journals/comsur/ZhaoLC17}. Therefore, to enhance the QoE of users, one should simultaneously consider these factors when conducting the I-UAV networking, which leads to a multi-objective decision problem. In this case, resorting to AI approaches will be a good choice to flexibly and dynamically learn the best I-UAV network control actions of maximizing the QoE according to the network environment.

%
%
%
%



\subsection{Situation aware Intelligent I-UAV networking}
UAVs are usually deployed in relatively low altitude 3D airspace and will share the airspace with other low altitude flying platforms. Owing the existence of some obstacles such as high-rise buildings, mountains, and high-tension lines, as shown in Figure 1, the low altitude airspace environment is complex. As a result, it is essential to consider live multi-domain situation on the entities in I-UAV networks and on the environment when conducting the I-UAV networking such that UAVs can complete missions safely and efficiently. 

The multi-domain situation is referred to as of evolution status of entities in time, frequency, spatial and network domains, including the UAV network status (e.g., network congestion condition, backbone nodes and links), UAV flight status, UAV safety and threat status.  

To guide the I-UAV networking with situation information, the perception of situation information and the dynamic construction of global situation field are required. 
However, the accurate situation perception and the construction of situation field are highly challenging. For instance, there are some dangerous small objects with unobvious characteristics (e.g. high-tension lines) in the low altitude airspace, as shown in Figure 1. How to accurately detect this type of objects is difficult. One will encounter the issues of data explosion, model explosion, and scene explosion when fusing the sensed situation, which challenges the construction of situation field. Besides, the intrinsic strong coupling feature of some flying objects in the time-spatial domain poses a challenge to the deduction of the global situation field. 

One can leverage some learning and game methods to tackle these challenge issues. For example, one can explore ML/AI approaches to effectively extract features of sensed objects or make some decisions based on the sensed objects. Next, leveraging AI approaches to realize the perception of situation information by fusing these features or decisions. 
A voting mechanism can be involved to identify the same object, which may be simultaneously sensed by multiple coordinated UAVs. A multi-agent dynamic evolution game strategy can be leveraged to constitute the global dynamic situation field from local perceived situation via situation coordination and deduction. 
The emerging meta learning method can also be explored to learn to fuse multi-modality sensor data. 


Additionally, constrained by the dynamic situation field, the solution of the I-UAV networking is non-trivial. The dynamic and high-dimensional constraints imply that one cannot leverage some conventional optimization methods and heuristic methods to mitigate the I-UAV networking problem. To this aim, some ML/AI approaches (e.g., Bayes network, reinforcement learning) should be considered to make good decisions by dynamically interacting with the situation field. It's noteworthy that the situation aware intelligent I-UAV networking has significant application value in both military and civilian fields. However, the research on it is still in its infant. 






\section{Networking of an airborne communication network: Challenges and Intelligent Approaches}
Compared to UAVs, HAPs have longer communication persistence, stronger service ability, and a larger footprint. For example, 16 HAPs with an elevation angle of 10$^{\circ}$ is enough to cover Japan, and Greece can be completely covered by eight HAPs \cite{DBLP:journals/comsur/KarapantazisP05}.
Meanwhile, compared to Satellites, HAPs are much closer to UAVs, which can significantly reduce the path loss and have shorter propagation delay and lower cost while faster responsiveness and higher flexibility. Therefore, it is essential to build a collaboratively integrated airborne communication network (ACN) consisting of I-UAV networks and HAP networks to significantly boost the communication coverage and enhance the service robustness. Certainly, these networks can work in conjunction with Satellite networks and terrestrial networks. 

The construction of a collaboratively integrated ACN, however, is highly challenging, where at least three critical issues should be tackled, i.e., channel estimation and tracking, integrated networking, and situation aware networking. 

From the perspective of channel estimation and tracking, the HAP-UAV wireless channel differs from the UAV-UAV channel due to unique propagation environment and transceiver antenna configuration. On one hand, the International Telecommunication Union Radio communication Sector (ITU-R) has allocated a number of frequency bands for HAP communications such as those in the Ka band (28-31 GHz and 47-48 GHz) \cite{grace2011broadband}. Signals in the Ka band are sensitive to time-varying atmospheric conditions such as oxygen, rain, and turbulence \cite{grace2011broadband}. 
The HAP-UAV wireless channel traverses the complex atmospheric propagation environment. Although many statistical HAP-ground channel models has been developed \cite{DBLP:journals/jsac/CaoYAXWY18}, they cannot effectively characterize the influence of time-varying climatic conditions on the channel modeling. To address this issue, on-line learning methods can be explored to estimate and track the HAP-UAV channel by approximating its path loss based on continuously measured channel coefficients. 
On the other hand, owing to the role of significantly increasing the HAP-UAV link capacity, a massive antenna array will be mounted on an HAP. The directivity and channel sparsity of massive antenna array, however, pose a great challenge to HAP-UAV channel estimation and tracking. Besides, owing to the long distance propagation, the slight disturbance in the flying posture of a HAP may cause the HAP-UAV pair to be misaligned. 
To tackle these issues, prediction joint with compress sensing methods should be investigated to estimate and track HAP-UAV channel. 

In terms of the integrated networking, HAP networks (in order to provide broadband services) and I-UAV networks may work at different frequency bands and have non-universal protocol stacks, which hinder the design of integrated ACN. Observing that it is unnecessary for each UAV to connect with an HAP. Thus, one can partition I-UAV networks into multiple I-UAV subnetworks (or called clusters) and explore learning methods to dynamically recommend a cluster head which will connect to an HAP. Next, one can utilize prediction methods to construct a dynamic time-spatial topology graph according to contact time and probabilities among HAPs and UAV cluster heads. With the dynamic topology graph in hand, unified protocols (e.g., routing protocols) can then be developed. 


In view of the situation aware ACN networking, it is more challenging than the situation aware I-UAV networking. In contrast to UAVs, an HAP can be deployed at an altitude between 17 km and 22 km, which indicates that the ACN network needs to construct a situation field of bigger and more complex airspace. In such airspace, the construction of situation field encounters more serious data, network and scene explosion issues. This is because more data (especially videos, images and radar data) in the airspace will be sensed and stored. To accurately analyze and learn these data, more complex and deeper NNs will be designed and trained. Meanwhile, the scenes are prone to sudden changes in such huge airspace, which poses a great challenge to the accurate sensing. 
Further, these explosion issues bring a strong pressure on the UAV and HAP computing power. UAVs and HAPs cannot realize efficient computing during a short period of time. In this case, how to prioritize data (e.g., identify small and critical data), fuse data, generate local situation information and deduce the global situation field by exploring intelligent approaches will be significant and crucial issues for in-depth research in the future.

\section{Conclusion}
An overview of the I-UAV networking was presented in this article. The basic network architecture of I-UAV and the intelligent modeling of channel gain of I-UAV networks were discussed. Moreover, challenges of QoS-driven, QoE-driven and situation aware I-UAV networking and the corresponding intelligent approaches to meet these challenges were highlighted. 
Nevertheless, some key technological breakthroughs regarding responsive and exact situation sensing and the global situation field construction should be made before the critical situation aware I-UAV networking can contribute to intelligent I-UAV networking standards. 
Lastly, potentials and challenges of future intelligent ACN networking were introduced. 
Further, we hoped that the discussion on the intelligent I-UAV and ACN networking in this article will inspire researchers' interest in the I-UAV and ACN networking and pave the way for them to design and build I-UAV and ACN networks in the future.


\ifCLASSOPTIONcaptionsoff
  \newpage
\fi




%
\bibliographystyle{IEEEtran}
\bibliography{networking}

%
\vspace{-12 mm}
\begin{IEEEbiographynophoto}{\text{ } \text{ } Peng Yang [S'17-M'19] (peng\_yang@buaa.edu.cn)}
received the Ph. D degree in Signal and Information Processing in 2018 from Beihang University. From 2019 to 2021, he was a Post-Doctoral Research Fellow with Singapore University of Technology and Design (SUTD), Singapore. Since 2021, he has been with Beihang University, where he is currently an Associate Professor. His current research topics include airborne communications and networking, network intelligence, URLLC, and airborne video transmission.

\textbf{Xianbin Cao [M'08$-$SM'10] (xbcao@buaa.edu.cn)}
is Dean and a Professor at the School of Electronic and Information Engineering, Beihang University, Beijing, China. His current research interests include intelligent transportation systems, airspace transportation management, and intelligent computation.
Currently, he serves as the Associate Editor of IEEE Transactions on Network Science and Engineering, and Associate Editor of Neurocomputing.

\textbf{Tony Q. S. Quek [S'98-M'08-SM'12-F'18] (tonyquek@sutd.edu.sg)}
received the B.E.\ and M.E.\ degrees in electrical and electronics engineering from the Tokyo Institute of Technology in 1998 and 2000, respectively, and the Ph.D.\ degree in electrical engineering and computer science from the Massachusetts Institute of Technology in 2008. Currently, he is the Cheng Tsang Man Chair Professor with Singapore University of Technology and Design (SUTD). He also serves as the Head of ISTD Pillar, Sector Lead of the SUTD AI Program, and the Deputy Director of the SUTD-ZJU IDEA. His current research topics include wireless communications and networking, network intelligence, internet-of-things, URLLC, and big data processing. 
He was honored with the 2008 Philip Yeo Prize for Outstanding Achievement in Research, the 2012 IEEE William R. Bennett Prize, the 2015 SUTD Outstanding Education Awards -- Excellence in Research, the 2016 IEEE Signal Processing Society Young Author Best Paper Award, the 2017 CTTC Early Achievement Award, the 2017 IEEE ComSoc AP Outstanding Paper Award, the 2020 IEEE Communications Society Young Author Best Paper Award, the 2020 IEEE Stephen O. Rice Prize, the 2020 Nokia Visiting Professor, and the 2016-2020 Clarivate Analytics Highly Cited Researcher. He is a Distinguished Lecturer of the IEEE Communications Society.

\textbf{Dapeng Oliver Wu [S'98$-$M'04$-$SM'06$-$F'13] (dpwu@ieee.org)}
received a B.E. degree in electrical engineering from Huazhong University of Science and Technology, Wuhan, China, in 1990, an M.E. degree in electrical engineering from Beijing University of Posts and Telecommunications, Beijing, China, in 1997, and a Ph.D. degree in electrical and computer engineering from Carnegie Mellon University, Pittsburgh, PA, in 2003.
Currently, he is a professor at the Department of Electrical and Computer Engineering, University of Florida, Gainesville, FL. His research interests are in the areas of networking, communications, signal processing, computer vision, machine learning, smart grid, and information and network security.
He was elected as a Distinguished Lecturer by IEEE Vehicular Technology Society in 2016. He has served as Chair for the Award Committee, and Chair of Mobile and wireless multimedia Interest Group (MobIG), Technical Committee on Multimedia Communications, IEEE Communications Society. He was an elected member of Multimedia Signal Processing Technical Committee, IEEE Signal Processing Society from Jan. 1, 2009 to Dec. 31, 2012.
\end{IEEEbiographynophoto}

%






\end{document}